\documentclass[preprint,prb,aps,superscriptaddress]{revtex4-1}  
\usepackage{graphicx}
\begin{document}       
%
%

\title{Room-temperature Low-field Colossal Magneto-resistance in Double-perovskite Manganite}
\author{S. Yamada$^{*}$}
\affiliation{Department of Materials System Science, Yokohama City University, Yokohama, 236-0027, Japan}
\author{N. Abe}
\affiliation{Department of Advanced Materials Science, The University of Tokyo, Kashiwa, 277-8561, Japan}
\author{H. Sagayama}
\affiliation{Institute of Materials Structure Science, High Energy Accelerator Research Organization, Tsukuba, Ibaraki 305-0801, Japan.}
\affiliation{Department of Materials Structure Science, The Graduate University for Advanced Studies, Tsukuba, Ibaraki 305-0801, Japan.}
\author{K. Ogawa}
\affiliation{Department of Materials System Science, Yokohama City University, Yokohama, 236-0027, Japan}
\author{T. Yamagami}
\affiliation{Department of Materials System Science, Yokohama City University, Yokohama, 236-0027, Japan}
\author{T. Arima}
\affiliation{Department of Advanced Materials Science, The University of Tokyo, Kashiwa, 277-8561, Japan}

\begin{abstract}    
The gigantic decrease of resistance by an applied magnetic field, which is often referred to as colossal magnetoresistance (CMR)\cite{Kus1,Cha1,Hel1,Jin1,Tokur1,Uru1,Tomi1,Tomi2,Tokur2}, has been an attracting phenomenon in strongly correlated electron systems. The discovery of CMR in manganese oxide compounds has developed the science of strong coupling among charge, orbital, and spin degrees of freedom. CMR is also attracting scientists from the viewpoint of possible applications to sensors, memories, and so on. However, no application using CMR effect has been achieved so far, partly because the CMR materials which satisfy all of the required conditions for the application, namely, high operating temperature, low operating magnetic field, and sharp resistive change, have not been discovered. Here we report a resistance change of more than two-orders of magnitude at a magnetic field lower than 2 T near 300 K in an A-site ordered NdBaMn$_{2}$O$_{6}$ crystal.  When temperature and a magnetic field sweep from insulating (metallic) phase to metallic (insulating) phase, the insulating (metallic) conduction changes to the metallic (insulating) conduction within 1 K and 0.5 T, respectively. The CMR is ascribed to the melting of the charge ordering. The entropy change which is estimated from the $B$-$T$ phase diagram is smaller than what is expected for the charge and orbital ordering. The suppression of the entropy change is attributable to the loss of the short range ferromagnetic fluctuation of Mn spin moments, which an important key of the high temperature and low magnetic field CMR effect. 
\end{abstract}

\maketitle

Colossal magnetoresistance (CMR)\cite{Kus1,Cha1,Hel1,Jin1,Tokur1,Uru1,Tomi1,Tomi2,Tokur2} in a series of perovskite manganites RE$_{1-x}$AE$_{x}$MnO$_{3}$ (where RE is a trivalent rare earth and AE a divalent alkaline earth) has attracted much attention for decades. The drastic phenomena are caused by the strong coupling among charge, orbital, and spin degrees of freedom. The negative magnetoresistance phenomena in perovskite manganites are classified into two categories. A fairly gradual and rather moderate negative magnetoresistance (MR) was observed around room temperature in La$_{1-x}$Sr$_{x}$MnO$_{3}$.\cite{Tokur1,Uru1} This reduction of resistance is attributed to the gradual increase in the band width by aligning the Mn spins in a magnetic field which is often referred to as double-exchange mechanism. As a result, the change in resistivity is also gradual and a high magnetic field of 5 T is needed to decrease the resistivity by one order. Another type of CMR effect, which is caused by the melting of charge ordering, was discovered in many compounds like Pr$_{1-x}$Ca$_{x}$MnO$_{3}$.\cite{Tomi1,Tomi2} The charge-order melting type CMR is much larger and steeper than that of the pure double-exchange type. For example, the resistivity in (Nd$_{0.06}$Sm$_{0.94}$)$_{1/2}$Sr$_{1/2}$MnO$_{3}$ at 115 K changes by more than two orders of magnitude in a low magnetic field below 1 T.\cite{Kuw1} The magnetic field necessary for a one-order decrease in resistivity is about 0.2 T. However, CMR of charge-order melting type in manganese oxide compounds has been rarely observed at room temperature so far, because the charge ordering phase transition temperature is lower than room temperature in most cases. Although some perovskite manganite compounds like Bi$_{1-x}$Sr$_{x}$MnO$_{3}$,\cite{Gar1,Gar2,Hejt1,Yam1} and REBaMn$_{2}$O$_{6}$ (RE = Sm - Y)\cite{Nak1,Ant1,Ari1,Tru1, Tro1, Nak2,Kag1,Aka1,Nak3,Aka2,Tru2,Nak4,Tru3,Tru4} exhibit the charge ordering phase above room temperature, such charge ordering phases are robust in a moderate magnetic field. In Sm$_{1-x}$La$_{x+y}$Ba$_{1-y}$Mn$_{2}$O$_{6}$,\cite{Nak4} which exhibits room-temperature CMR effect by melting the charge ordering phase, for example, a magnetic field of 9 T is necessary for the resistivity change of one order of magnitude. In this paper, we report room-temperature CMR in a NdBaMn$_{2}$O$_{6}$ single crystal. Previous studies using polycrystalline samples revealed that the magnetic susceptibility suddenly dropped with cooling across 290 K. This anomaly was attributed to the A-type (layered) antiferromagnetic ordering of Mn spin moments.\cite{Nak2,Aka1,Aka2} We recently succeeded in single-crystal growth of NdBaMn$_{2}$O$_{6}$ and proposed that a first-order metal-insulator phase transition should not be accompanied by any  magnetic ordering.\cite{Yam2} The resistivity change upon the metal-insulator transition is still larger than two orders of magnitude. The metal-insulator transition temperature $T_{MI}$ is decreased by applying a magnetic field. A high-quality single crystal exhibits the metal insulator transition near 300 K. When temperature sweeps from insulating (metallic) phase to metallic (insulating) phase, the insulating (metallic) conduction changes to the metallic (insulating) conduction within 1 K. The crystal exhibits a room-temperature CMR effect at a relatively low field of 2 T.

\begin{figure} 
\includegraphics[width=10cm]{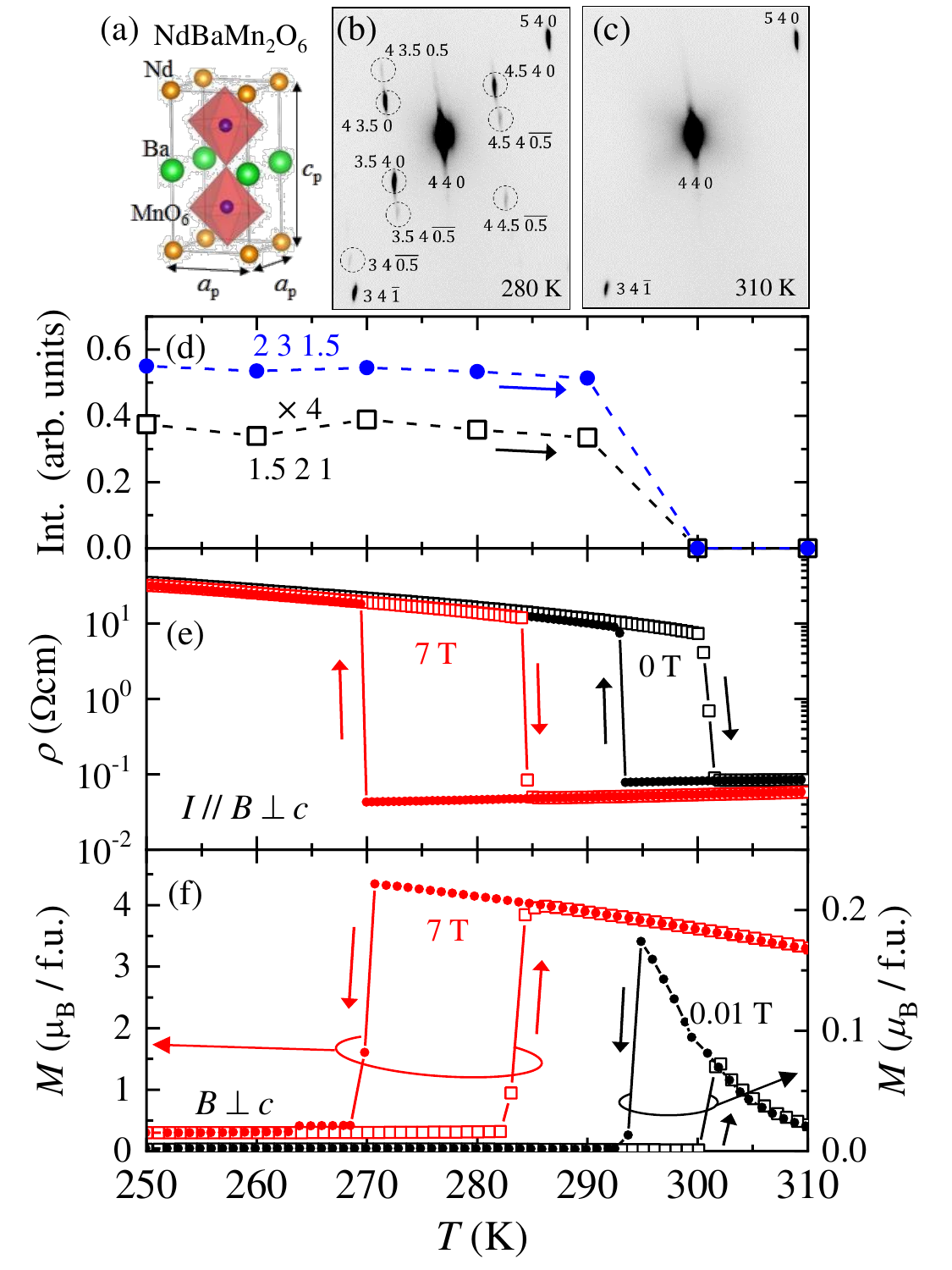} 
\caption{X-ray oscillation photographs of a NdBaMn$_{2}$O$_{6}$ single crystal around 4 4 0 at (b) 280 K and (c) 310 K.  The indices are given for the $\sqrt{2} a_{p} \times \sqrt{2} a_{p} \times c_{p}$ unit cell, where $a_{p}$ and $c_{p}$ are lattice constants of simple tetragonal unit cell shown in (a). Temperature dependence of (d) integrated intensities of 2 3 1.5 and 1.5 2 1 superreflections, (e) electrical resistivity at 0 T, and at 7 T, and (f) magnetization at 0.01 T and 7 T. The magnetic field is applied perpendicular to the $c$-axis, and parallel to the electric current in (e).}
\label{f:1}
\end{figure}

@Figure \ref{f:1} shows temperature dependence of intensities of x-ray superlattice reflections, electrical resistivity, and magnetization of NdBaMn$_{2}$O$_{6}$ single crystal. The magnetization and resistivity show an anomaly near $T_{MI}=$ 300 K in the warming run. In a magnetic field of 7 T, the phase transition temperature in the warming run decreases to 283 K. The insulating phase between 283 K and 300 K hence changes to the metallic phase by the application of a magnetic field. The resistivity change is more than two orders of magnitude. Because the resistivity does not depend on temperature very much either in the metallic or insulating phase, the magnitude of CMR is not sensitive to temperature. Though the phase transition is accompanied with a hysteresis, the resistivity change is very sharp. When temperature sweeps from insulating (metallic) phase to metallic (insulating) phase, the insulating (metallic) conduction changes to the insulating (metallic) conduction within 1 K. It is of note that the metallic and insulating phases coexist in the hysteresis. The metal-insulator transition is accompanied with a magnetic anomaly, as shown in Fig. \ref{f:1}(f). Though the magnetization above $T_{MI}$ increases with a magnetic field, the magnetization below $T_{MI}$ does not depend on a magnetic field very much. Single-crystal x-ray oscillation photographs indicate the charge and orbital ordering in the insulator phase below $T_{MI}$. All the x-ray reflections in this paper are indexed on the basis of the $\sqrt{2} a_{p} \times \sqrt{2} a_{p} \times c_{p}$ unit cell, where $a_{p}$ and $c_{p}$ are lattice constants of simple tetragonal unit cell, as shown in Fig. \ref{f:1}(a). The crystal contains $ab$ twin structure as mentioned in our previous paper.\cite{Yam2} Only the fundamental reflections of the $\sqrt{2} a_{p} \times \sqrt{2} a_{p} \times c_{p}$ unit cell  are observed above $T_{MI}$ as shown in Fig. \ref{f:1}(c). In contrast, many superlattice reflections appear around 4 4 0 reflection below $T_{MI}$ in Fig. \ref{f:1}(b). The superlattice reflections indicate that the unit cell below $T_{MI}$ is $2\sqrt{2} a_{p} \times \sqrt{2} a_{p} \times 2 c_{p}$. This unit cell is the same as that of charge ordering phase of SmBaMn$_{2}$O$_{6}$ between 200 K and 380 K.\cite{Sag1} Here it is of note that the intensities of the superlattice reflections are more than six orders weaker than some fundamental reflections. In our previous study\cite{Yam2}, we used a crystal of a dimension 50 $\mu$m $\times$ 10 $\mu$m  $\times$ 10 $\mu$m and could not detect superlattice reflections. In the present x-ray study we used a much larger crystal with a dimension of 1 mm $\times$ 2 mm  $\times$ 2 mm. Then we could observe a series of superlattice reflections.\cite{Yam5} Some previous studies by means of neutron powder diffraction measurements reported that Mn moments should be aligned to form the A-type antiferromagnetic order.\cite{Nak2}  The checkerboard-type charge ordering is obviously inconsistent with the A-type antiferromagnetic ordering. The disagreement might be caused by some off-stoichiometry or A-site intermixing. A further neutron study would settle the problem. 

\begin{figure} 
\includegraphics[width=10cm]{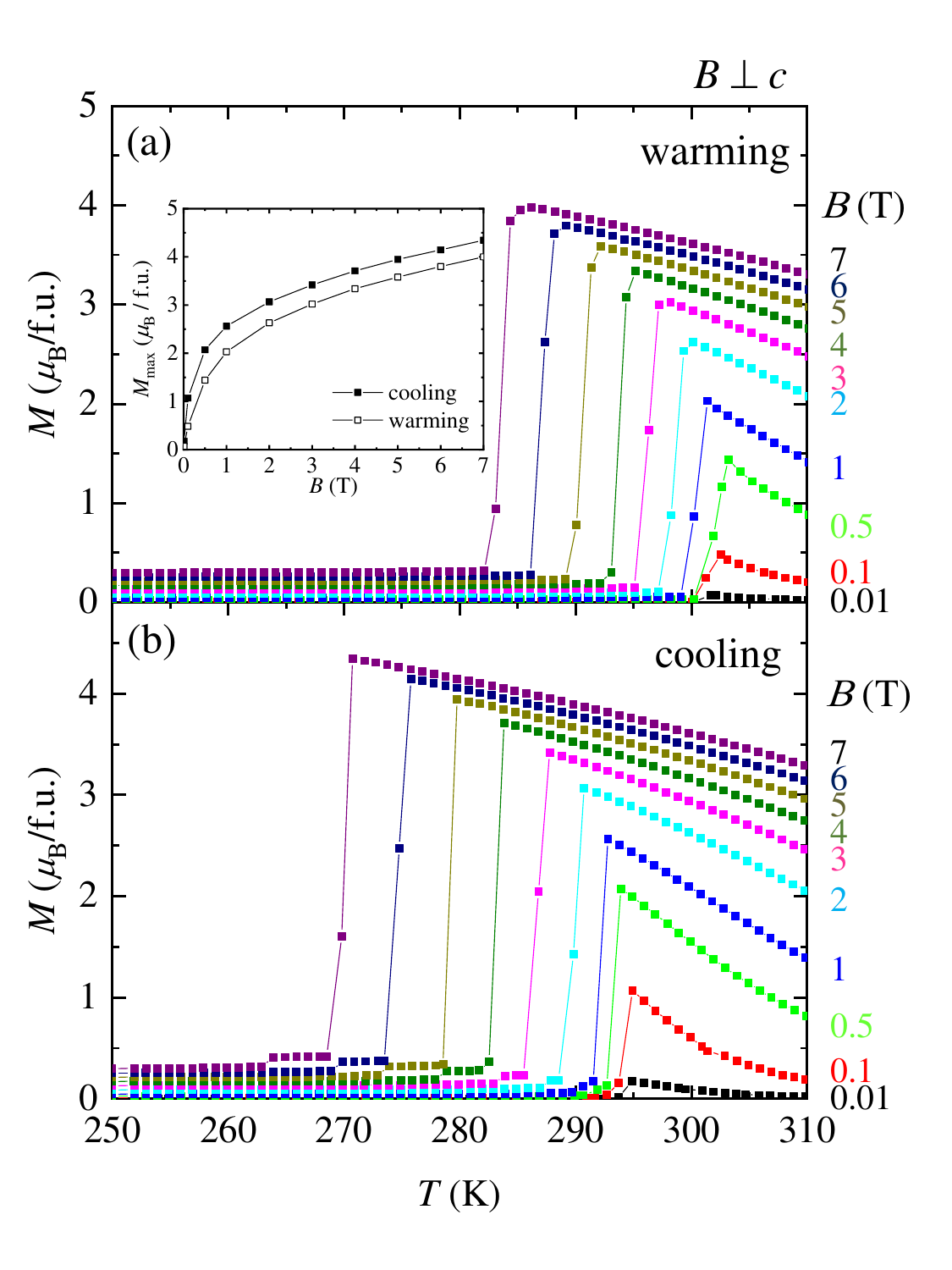} 
\caption{Temperature dependence of magnetization in (a) warming  and (b) cooling runs. The magnetic field is applied perpendicular to the $c$ axis. Inset show the maximum of the curve as a function of the applied magnetic field.}
\label{f:2}
\end{figure}

Temperature dependence of magnetization at several magnetic fields is shown in Fig. \ref{f:2}. $T_{MI}$ monotonically decreases with a magnetic field. We define the peak value of the magnetization, which is present at  just above $T_{MI}$ as $M_{max}$. $M_{max}$ steeply increases with an applied magnetic field, as shown in an inset of Fig. \ref{f:2}. $M_{max}$  in the cooling run at 7 T, 4.3 $\mu_{B}$/f.u., is much smaller than 7 $\mu_{B}$/f.u., the simple estimation of the fully saturated moment. 

\begin{figure} 
\includegraphics[width=10cm]{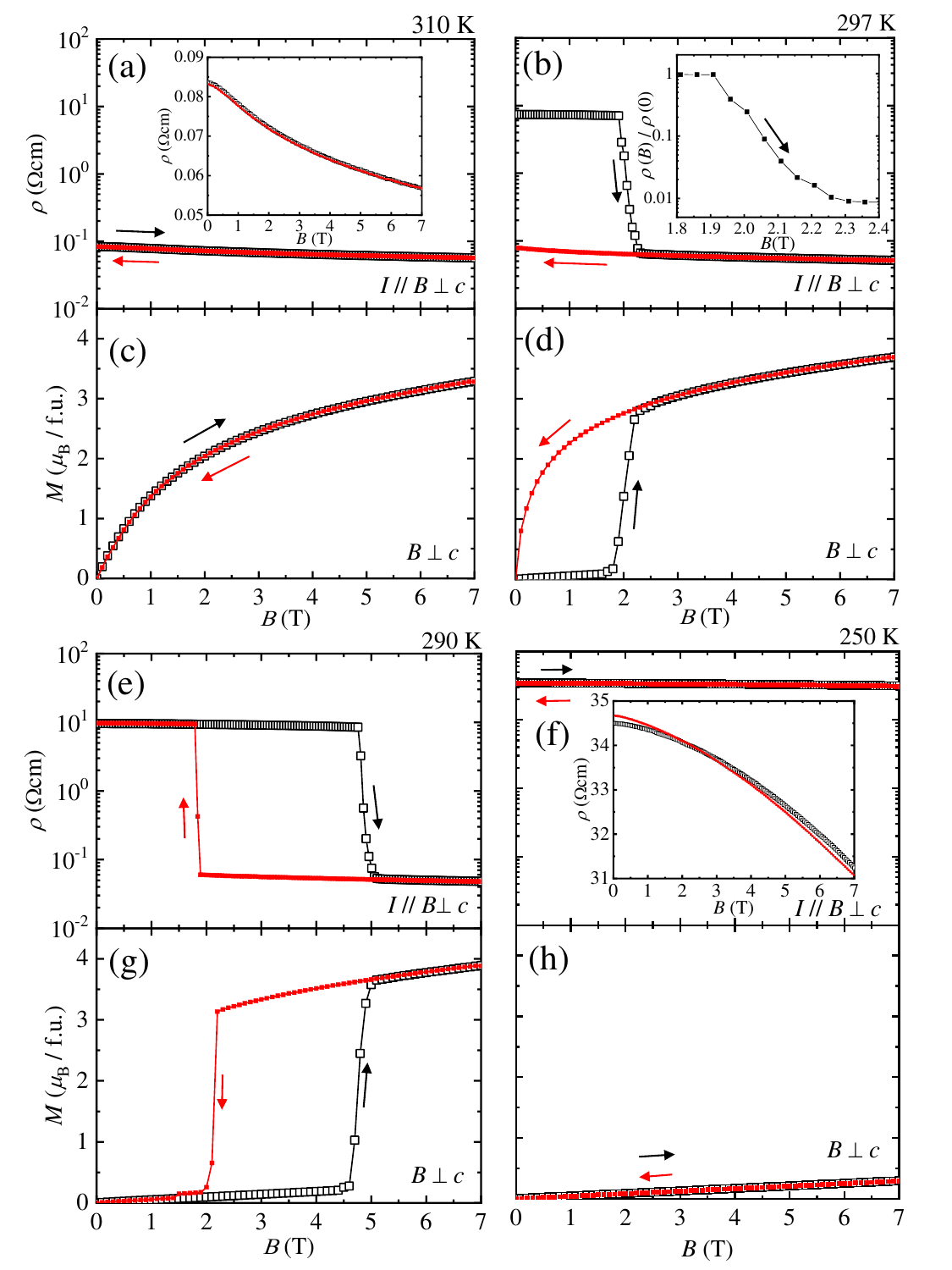} 
\caption{Magnetic field dependence of (a,b,e,f) resistivity and (c,d,g,h) magnetization at several temperatures. Insets in (a) and (f) are (a) and (f) whose vertical scales are linear, respectively. Inset in (b) is enlarged figure in the increasing run of a magnetic field in (b) around the metal-insulator transition. The resistivity of the inset in (b) is normalized by the resistivity at zero magnetic field.(see text) }
\label{f:3}
\end{figure}

Magnetic field dependence of resistivity and magnetization around $T_{MI}$ is shown in Fig. \ref{f:3}. At 310 K, which is just above $T_{MI}$, the resistivity slightly decreases with an increase of the magnetic field, while the magnetization steeply increases in a low magnetic-field range. These results imply the presence of ferromagnetic fluctuation above $T_{MI}$, as reported in our previous paper.\cite{Yam3} The relative resistivity change at a magnetic field $B$ is often defined as  
\begin{equation}
\frac{\rho(0)-\rho(B)}{\rho(B)}.
\label{eq:1}
\end{equation}
This value is as small as 35 \% at 7 T, 310 K. At 297 K, which is just below $T_{MI}$, a metamagnetic transition which is accompanied with a metal-insulator phase transition is clearly observed. The relative resistivity change at 7 T exceeds 14000\%. This value is ten times as large as that of Sm$_{1-x}$La$_{x+y}$Ba$_{1-y}$Mn$_{2}$O$_{6}$ at 9 T at 300 K.\cite{Nak4} The resistivity change is very sharp. For example, at 297 K around the phase transition, the resistivity starts decreasing steeply at 1.9 T in a field increasing run, and the value of the resistivity at 2.1 T is more than one order of magnitude smaller than that at 1.9 T as shown in the inset of Fig. \ref{f:3}(b). The steep change is almost finished at 2.4 T. The resistivity change with sweeping the magnetic field is as steep as that in (Nd$_{0.06}$Sm$_{0.94}$)$_{1/2}$Sr$_{1/2}$MnO$_{3}$ at 115 K.\cite{Kuw1} The phase-transition field increases with cooling. At 297 K, the insulator phase never revives just by sweeping the external magnetic field, if the field-induced insulator-to-metal transition once takes place. In contrast, at 290 K, the magnetic-field induced insulator-metal transition becomes reversible. The steepness of the resistivity change in a magnetic field decreasing run is almost the same as that in a magnetic field increasing run. At 250 K, the insulator-metal transition is not observed up to 7T and the relative resistivity change at 7 T is as small as 10\%.

\begin{figure} 
\includegraphics[width=10cm]{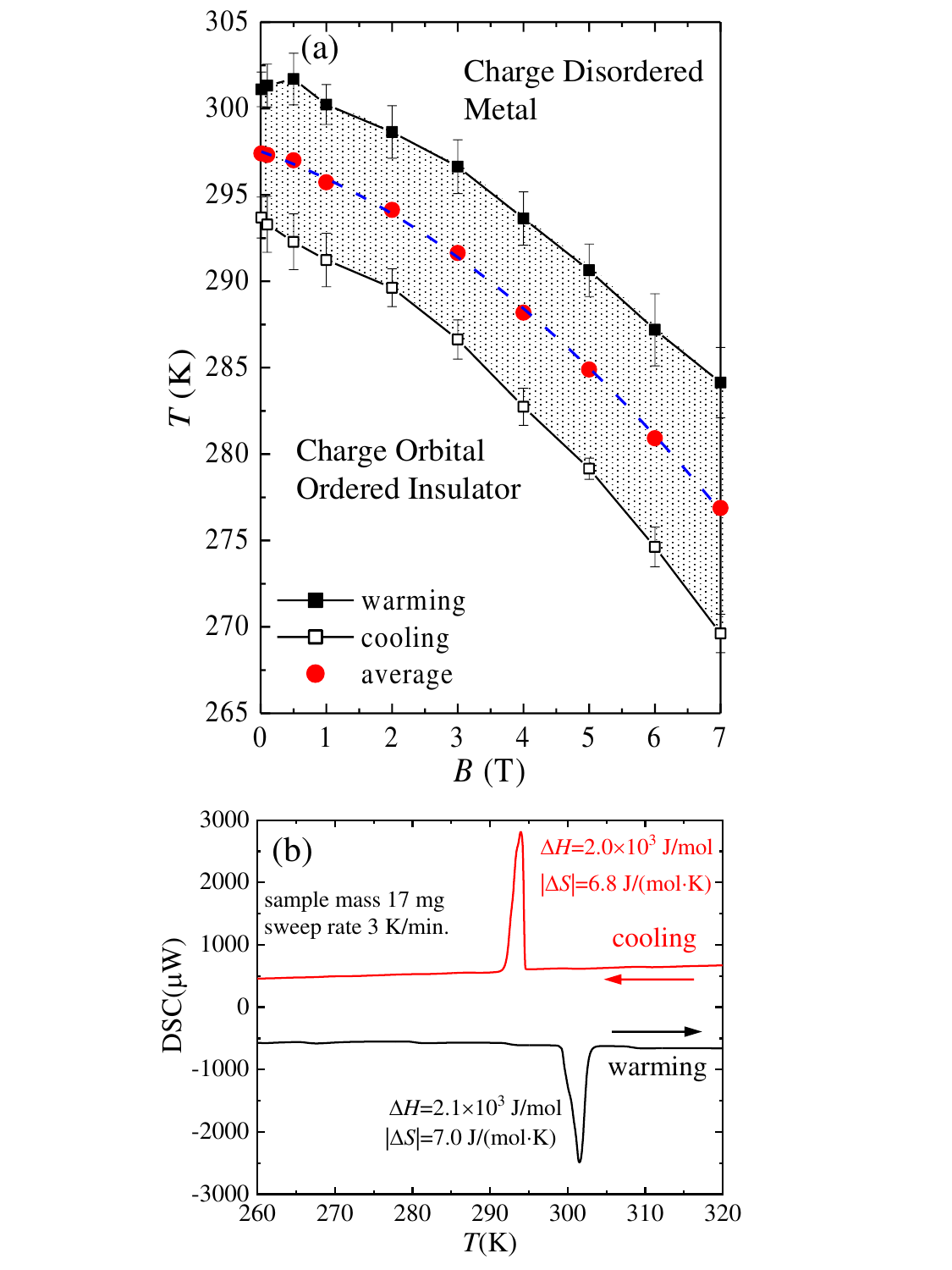} 
\caption{(a) Phase diagram of NdBaMn$_{2}$O$_{6}$ in the magnetic field temperature plane. Red solid circles show the average values of $T_{MI}$'s for cooling (open squares) and warming (solid squares) runs. Blue broken line is a fitting curve to the magnetic field dependence of the averaged $T_{MI}$.(see text) (b) Differential scanning calorimetry around $T_{MI}$.}
\label{f:4}
\end{figure}

Next, we discuss the necessary condition for the low-field CMR at a wide temperature range. Clausius-Clapeyron equation for a first-order magnetic phase transition is given as 
\begin{equation}
\frac{dT_{c}}{dB} = -\frac{\Delta M}{\Delta S}, 
\label{eq:2}
\end{equation}
where $T_{c}$, $\Delta M$, and $\Delta S$ are the phase transition temperature, difference in magnetization between two phases, and the entropy change upon the phase transition, respectively. Fig. \ref{f:4}(a) shows the phase diagram on the magnetic-field-temperature plane. Because the present metamagnetic transition is accompanied with a hysteresis, the thermally equilibrium phase boundary cannot exactly be determined experimentally. Here we assume that thermally equilibrium $T_{c}$ is as high as average of the experimentally obtained $T_{MI}$'s in the warming and cooling runs at a constant magnetic field. The magnetic field dependence of $T_{MI}$ is fitted by a quadratic function of a magnetic field $B$ (a blue broken line in Fig. \ref{f:4}).\cite{fit}  Then we calculate $\Delta S$ from eq.(\ref{eq:2}). The $\Delta S$ values are $4 \sim 7$ J/(mol$\cdot$K) above 1T.\cite{fit2} $\Delta S$ at zero magnetic field is calculated from the latent heat, shown in Fig \ref{f:4}(b), to be about 7 J/(mol$\cdot$K). In the charge disordering phase, the entropy in the charge and orbital sectors in the doubly degenerate $e_{g}$ orbitals is roughly estimated as $(3N_{{\rm Mn}}/2)k_{B}\ln2 \simeq 17.3$ J/(mol$\cdot$K), where $N_{{\rm Mn}}$ and $k_{B}$ are the number of Mn atoms in one molar NdBaMn$_{2}$O$_{6}$ and Boltzmann constant, respectively. The entropy in the charge and orbital sectors is about 10  J/(mol$\cdot$K) larger than the measured value of the entropy change. The change in Mn spin states should also attribute to the entropy change. Above $T_{MI}$, the short-range ferromagnetic contribution is enhanced with approaching $T_{MI}$, and suddenly disappears below $T_{MI}$. The Mn spin moments remain paramagnetic between N\'eel temperature (235 K) and $T_{MI}$.\cite{Yam2} In other words, the entropy in the spin sector increases with cooling across $T_{MI}$. In the paramagnetic phase, the spin entropy is estimated to be $(N_{{\rm Mn}}/2)k_{B}\ln(5\times4) \simeq 24.9$ J/(mol$\cdot$K) by neglecting the Mn spin correlation. If we neglect the change of lattice entropy upon the phase transition, the spin entropy above $T_{MI}$ is roughly calculated to be about 15 J/(mol$\cdot$K). This value could be explained by assuming the formation of ferromagnetic clusters of 7 $\sim$ 8 Mn$^{3.5+}$ ions above $T_{MI}$. The low magnetic-field CMR of NdBaMn$_{2}$O$_{6}$ should be closely related to the competition between the charge and orbital sectors and the spin sectors. The CMR effect in a low magnetic field in a wider temperature range can be realized by a further increase in  $\Delta M / \Delta S$. If the ferromagnetic cluster above $T_{MI}$ becomes larger, it is expected that the operating magnetic field of the CMR effect becomes lower in a wide temperature range because both of the enhancement of $\Delta M$ and the suppression of $\Delta S$ are satisfied simultaneously. 

The above mentioned discussion suggests the necessary conditions for the high-temperature low-field CMR effect in the manganese oxide compounds. First, the charge-orbital ordering temperature  T$_{COO}$ must be high, because the charge-order melting type CMR can only appear below $T_{COO}$. Second, ferromagnetic ordering or fluctuation must be present above $T_{COO}$ for large $\Delta M$ . Third, the charge-orbital order phase just below $T_{COO}$ must not be accompanied by antiferromagnetic ordering for a partial cancellation of a large entropy change in the charge and orbital sectors. In general, the first and second conditions are difficult to satisfy simultaneously. When the Mn-O-Mn bond angle is close to 180 degrees, the ferromagnetic and metallic behavior becomes dominant due to the large double exchange interaction. When the bond angle becomes smaller, antiferromagnetic and insulating behavior becomes dominant because of the suppression of the transfer of $e_{g}$ electron. In other words, if the charge ordering phase is stable up to high temperatures, the ferromagnetic interaction would not survive above the charge ordering phase transition temperature. However, in NdBaMn$_{2}$O$_{6}$, the charge ordering transition temperature reaches room temperature, while the NdBaMn$_{2}$O$_{6}$ has a large instability of ferromagnetism. In double perovskite manganites, random potential, which is harmful for any ordering, can be minimized. Furthermore, the average valence of Mn ions can be set to 3.5 exactly, which would favor the charge orbital ordering. The anisotropic crystal structure may also assist the formation of orbital ordering. As a result, the charge-orbital order can be stabilized even if the $e_{g}$ electron transfer energy is fairly large. Here the alternate arrangement of RE and Ba would be essential for the reduction of the random potential. Nakajima et al.\cite{Nak3} showed that the enhancement of ordering ratio between RE and Ba atoms should raise $T_{MI}$ and sharpen the transition. The intersite mixing between Nd and Ba in a single crystal with $T_{MI}$ of 290 K in the warming run is estimated  the to be 1.2 \% by using single crystal structure analysis by means of the synchrotron x-ray diffraction measurements.\cite{Yam2} In the present study, $T_{MI}$ is higher than 290 K, suggesting a highly ordered A-site cations. Although the metal-insulator transition is accompanied with a large hysteresis for temperature and a magnetic field dependence due to the first-order nature of the transition, the resistivity changes for temperature and a magnetic field sweeping are very steep. This result means that the spatial randomness due to the intersite mixing between Nd and Ba strongly affects not only the stability of the charge and orbital ordering but also the steepness of the resistivity change with the phase transition. Once some nucleation happens in a crystal, the embryos  can grow rapidly without heavy pinning of domain walls. Moreover the resistivity change of the single crystal at the phase transition is steeper than that of polycrystalline sample\cite{Aka1} due to the absence of the grain boundary resistance. The steepness of the resistivity change at the phase transition in the NdBaMn$_{2}$O$_{6}$ single crystals used in the present study would propose the almost complete removal of randomness. 

In conclusion, colossal magnetoresistance (CMR) than two-orders of magnitude at a magnetic field lower than 2 T has been observed near 300 K in NdBaMn$_{2}$O$_{6}$ single crystal. The magnitude does not depend on temperature very much. Single crystal synchrotron x-ray diffraction has revealed that the insulating state below 300 K should be ascribed to the charge and orbital ordering. The charge ordering phase near room-temperature is melted by a relatively low magnetic field. Although the phase transition is accompanied with a large hysteresis, the resistivity steeply change in temperature and magnetic field sweeping. Such the steepness of the resistivity change at phase transition could be attributed to the almost complete removal of randomness by a highly ordered Nd and Ba atoms. 

\section{METHODS} 
  Single crystals of NdBaMn$_{2}$O$_{6}$ were grown by a floating zone (FZ) method. Powders of Nd$_{2}$O$_{3}$, BaCO$_{3}$, and Mn$_{3}$O$_{4}$ were mixed, ground, and calcined at 1290 $^\circ$C for 48 hours in Ar (6N) atmosphere. The resultant powder was pulverized, shaped into a cylinder under a hydrostatic pressure of 30 MPa and sintered at 1290 $^\circ$C for 12 hours in Ar (6N) atmosphere to form feeding and seeding rods. An FZ furnace equipped with two halogen incandescent lamps and hemielliptic focusing mirrors was used for the crystal growth. The molten zone was vertically scanned at a rate of 2 mm/h in Ar gas mixed with a tiny portion ($\leq 0.1$ \%) of H$_{2}$. The melt-grown bar was annealed in O$_{2}$ atmosphere at 500 $^\circ$C for 48 hours. Magnetization was measured using a commercial superconducting quantum interface device magnetometer (Quantum Design MPML-XL). Electrical resistivity and magneto resistance were measured using a commercial system (Quantum Design PPMS) by the conventional four-probe method. For measuring the magnetic field dependence, the sample was warmed from 250 K to the target temperatures at 0 T. Single-crystal synchrotron radiation x-ray diffraction measurement was performed at Photon Factory, High Energy Accelerator Research Organization (KEK), Japan. The dimensions of the single crystal sample were 1 mm $\times$ 2 mm $\times$  2 mm. X-ray-diffraction intensities were collected by the oscillation photograph method with a cylindrical imaging-plate diffractometer at BL-8B. The collected data were analyzed by RAPID AUTO (Rigaku Co.).The temperature was controlled by using a He-gas spray-type refrigerator. The wavelengths were selected to be 0.688 ${\rm \AA}$. Latent heat of a NdBaMn$_{2}$O$_{6}$ single crystal of 17 mg was measured using a commercial differential scanning calorimeter (Hitach Hightech DSC-7020) with a reference sample of Al$_{2}$O$_{3}$. The sweeping rate of temperature was set 3 K/min.

\acknowledgments
This study was partly supported by a Grant-in-Aid for Scientific Research (No. 1818K03546) from the Japan Society for the Promotion of Science and by the grant for Strategic Research Promotion(No.G2503) of Yokohama City University. This work was performed under the approval of the Photon Factory Program Advisory Committee (Proposal No. 2014G597). Magnetoresistance measurements were performed using facilities of the Institute for Solid State Physics, the University of Tokyo.


\begin{thebibliography}{99}
\bibitem{Yam5}Some fundamental reflections were so strong that the data reached the saturation limit.
\bibitem{fit} The function of the fitting curve is $T(B)= -0.233B^{2}-1.34B+296$.
\bibitem{fit2} We also tried to fit the phase transition curve using an even and a quadratic function. However, the fitting curve does not represent the phase transition curve in both low and high magnetic field ranges. The function could be a higher order even function.
\end{thebibliography}
\end{document}